# Skynet's "Astrophotography of the Multi-Wavelength Universe!" Curriculum

Daniel E. Reichart

*Department of Physics and Astronomy, University of North Carolina at Chapel Hill, Chapel Hill, NC 27599-3255*

**Abstract.** Here we briefly describe each of the modules that constitute Skynet's new "Astrophotography of the Multi-Wavelength Universe!", or MWU!, curriculum.

## 1.   Skynet Curricula and MWU!

Over the past two decades, UNC-Chapel Hill has built one of the largest networks of fully automated, or robotic, telescopes in the world, significantly advancing this new technology.  These telescopes are used both for cutting-edge research and for cutting-edge education:  Funded by a series of large NSF awards, (1) we have developed unique, student-level, observing and image-analysis interfaces, allowing students, of all ages, to use this globally distributed, research tool, right alongside the professionals; and (2) often in partnerships with professional educators and education researchers, we have developed a sequence of observation-based curricula and experiences that leverage these hardware and software resources, from the elementary-school level through the graduate-school level, reinforcing and strengthening the STEM pipeline.

Now funded by a $3M DoD STEM award, we are integrating a global network of 10m – 30m diameter radio telescopes into Skynet, and developing a follow-up to curriculum to our popular, survey-level "Our Place in Space!", or OPIS!, curriculum [1,2].  The new curriculum is called "Astrophotography of the Multi-Wavelength Universe!", or MWU! [3,4].  MWU! consists of observing experiences that integrate optical *and* radio, as well as archival infrared data, and the topics of inquiry span solar system objects; star formation, evolution, and death; normal and active galaxies and their evolution; and light-producing mechanisms.  Astrophotography serves as this curriculum's "hook".

In this document, we briefly describe each of the modules, current versions of which are linked to in this example syllabus:  https://tinyurl.com/mwu-syllabus.

## 2.   Module 1.  The Solar System at Visible and Invisible Wavelengths

The goal of Module 1 is to use bright (i.e., low-cost, in terms of telescope time) solar-system targets (1) to ramp students up on MWU!'s key technologies and concepts, and (2) to introduce them to radio astronomy on an equal footing with optical astronomy. Module 1 consists of four separate investigations.

## 2.1. Module 1A. The Moon at Optical Wavelengths: Mosaicking and Mineralogy

Students learn basic color combination and mosaicking, using the moon. Science content (color-based lunar mineralogy) is low, but it is a fun / visually exciting way to quickly build up Skynet and Afterglow skill sets (see Figure 1).

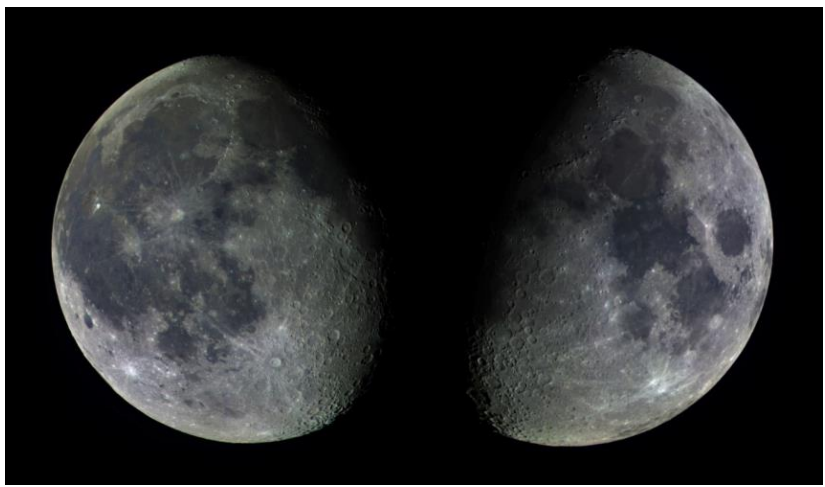

Figure 1. Two mosaics of the moon. Each consists of 5 × 5 = 25 tiles × 3 filters (red, green, blue) = 75 images. The colors have been saturated to highlight different aspects of lunar mineralogy.

## 2.2. Module 1B. The Moon at Radio Wavelengths: Reflection vs. Blackbody Emission / Module 1C. Jupiter at Radio Wavelengths: Blackbody vs. Synchrotron Emission

Students pick either Module 1B (moon) or Module 1C (Jupiter), and learn radio mapping, aperture photometry, and photometric calibration (see Figure 2). All three are core skills/concepts. They also engage in hypothesis testing (reflected sunlight vs. thermal moonlight for Module 1B, and thermal vs. non-thermal emission for Module 1C). In both cases, they measure temperature in the Rayleigh-Jeans limit.

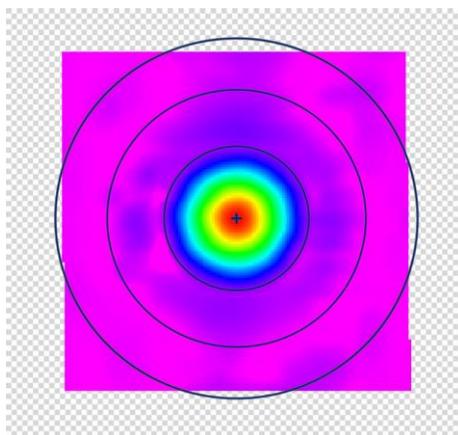

Figure 2. An artificially-colored image of the moon, made using Skynet's Radio Cartographer codebase [5]. Overlayed is an aperture, for measuring the brightness of the source, and an annulus, for measuring the background level. Between the aperture and the annulus, one can see a faint, Airy diffraction ring, which can be seen around bright sources in well-focused telescopes, radio or optical.

## 2.3. Module 1D. The Planets at Optical Wavelengths: Percentile vs. Photometric Color Balance

Students pick up a variety of key skills, including observing and stacking images for depth, aligning images using only a single source, and adding (world) coordinate systems (WCS) to images. They also learn the difference between percentile color balance, which is fast, easy, and technically/often wrong, and photometrically calibrated color balance, which is used through the rest of MWU!. Science content includes surface and cloud-top chemistry/composition (see Figure 3).

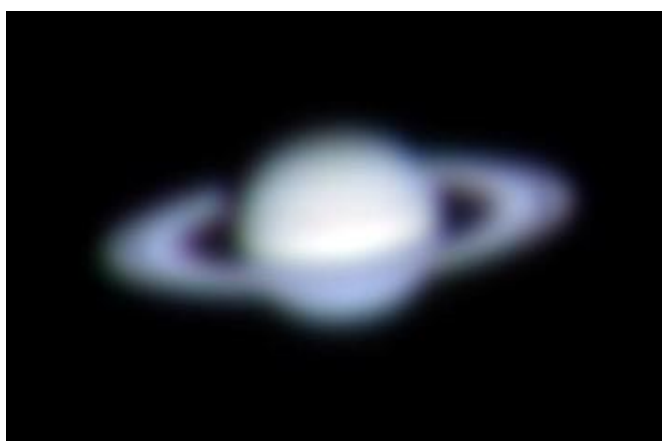

Figure 3. A quick, color composition of Saturn, made and de-pixelated in Skynet's Afterglow image-processing application. Butterscotch indicates ammonia ice. White indicates water ice.

### 3.    Module 2.  Star Clusters:  HR Diagrams and Stellar Evolution

Students observe young, intermediate-age, and old star clusters (two open clusters and a globular cluster).  They learn (1) how to plot HR diagrams, (2) how to use Gaia data to clean these diagrams and measure a cluster's distance, and (3) how to match isochrones to these diagrams to measure a cluster's age, metallicity, and reddening (see Figure 4).  In doing so, they learn how stars evolve from the main sequence to the red giant branch to the horizontal branch.  They make reddening-corrected images and make associations between stellar colors and brightnesses in their images and components in their HR diagrams.  In an inter-team activity, they learn to guesstimate star-cluster ages from reddening-corrected images alone.

Much of this module can also be carried out without access to Skynet telescopes, and consequently has the potential to be used very broadly [6].

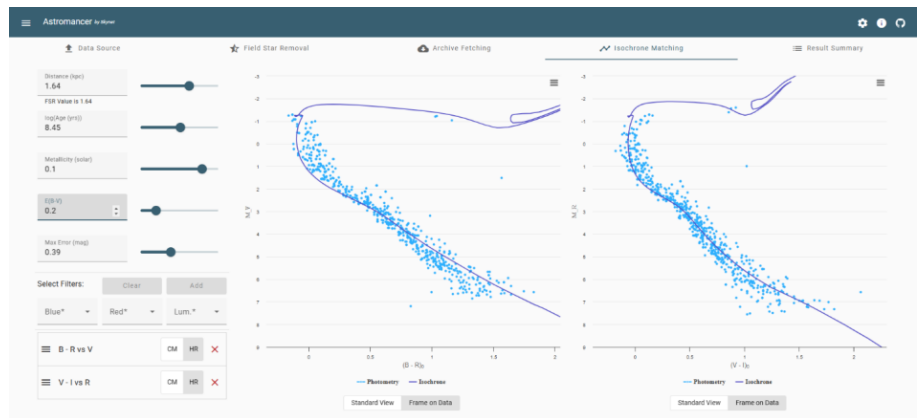

Figure 4.  A screenshot from Clustermancer, where the user has by-eye matched isochrones to brightness vs. color information from Skynet-acquired images of each star in intermediate-age, open cluster NGC 2437, dating the cluster, and measuring its metallicity and line-of-sight reddening due to dust.

### 4.    Module 3.  Introduction to Astrophotography:  Star Birth

Students select a (currently observable) star-forming region (SFR) for an in-depth study, and for their first serious astrophotography project.  They learn how to plan an observation, and carry out a test observation to determine optimal exposure durations and viable narrowband (NB) filters.  Then, they carry out a deep, LRGB+NB observation.  They also carry out a small radio map of the region, looking for bremsstrahlung emission.  They learn how to use a luminance (L) layer to increase signal-to-noise, narrowband layers to highlight excited, star-heated gas, and archival mid-infrared (MIR) and near-infrared (NIR) layers to reveal star-heated dust and dust-hidden stars, respectively (see Figure 5).  Finally, combining information from all of their image's layers, they carry out Strömgren sphere and pressure-balance calculations, measuring the SFR's density, and placing lower and upper limits on the surrounding cloud's density [7].

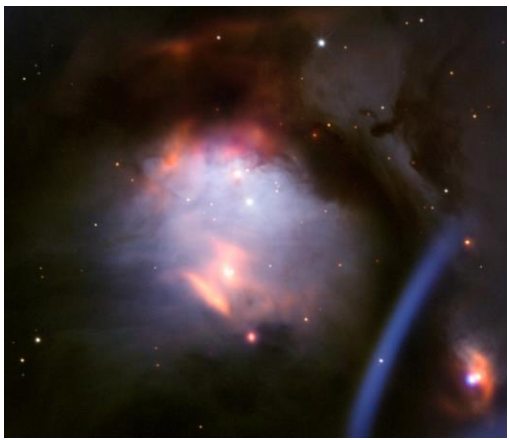

Figure 5. LRGB image of reflection nebula Messier 78, acquired using Skynet's 16-inch diameter PROMPT-6 telescope at Cerro Tololo Inter-American Observatory (CTIO) in Chile. This has been supplemented with 2MASS NIR data at 2 microns to reveal dust-hidden stars, and NASA WISE MIR data at 12 and 22 microns to reveal star-heated dust.

## 5. Module 4. The Invisible Universe

Although students learn core radio astronomy skills and concepts in Module 1B and/or 1C, this is their first serious exploration of the invisible, radio sky.

### 5.1. Module 4A. Radio Survey and Source Catalog

We divide the plane of the Milky Way into 15 ≈400 square degree regions. Each student team maps at least one of these regions, and "discovers" star-forming regions, supernova remnants, and supermassive black holes contained within. The class then combines their smaller maps into a larger map (see Figure 6), and compiles a catalog of discovered sources and their measured brightnesses.

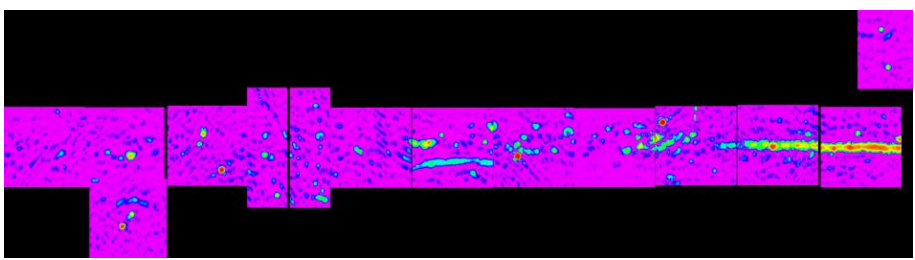

Figure 6. Map of the plane of our Milky Way galaxy at radio wavelengths, made with Skynet's 20m-diameter radio telescope at Green Bank Observatory (GBO) in West Virginia, using [5].

## 5.2. Modules 4B. Color at Radio Wavelengths: Emission Mechanisms

Using small maps of the brightest radio sources, some of which they have already collected for previous modules, students learn how to, post facto, split their data into three (RGB) sub-bands, and using the resulting maps, how to make calibrated and "naturalized" – meaning as our eyes would see it if they worked at radio wavelengths – color images. Thermal sources have a +2 spectral index, and consequently appear "thermal" blue, like O stars do to the eye at optical wavelengths. Bremsstrahlung sources (e.g., star-forming regions) appear blue to white. Synchrotron sources (e.g., supernova remnants and supermassive black holes at the cores of active galaxies) appear white to light red. In this way, emission mechanisms can be identified by color (see Figure 7). Students develop a deeper understanding of these continuum-emitting processes, and learn a new color vocabulary, applicable to the invisible universe.

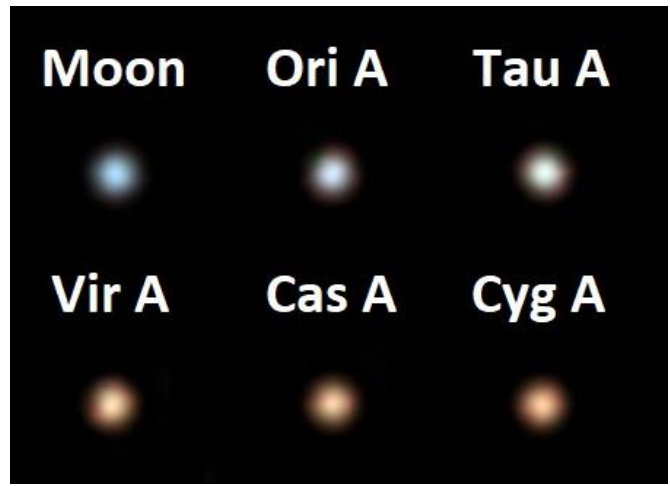

Figure 7. Naturalized color images of the six brightest radio sources in the sky (other than the sun), made with Skynet's 20m-diameter radio telescope at GBO. At optical wavelengths, the color of a source usually indicates its temperature. Here, color indicates the light's emission mechanism (thermal – blue, bremsstrahlung – blue to white, or synchrotron – white to light red).

## 6.    Module 5. Star Death: White Dwarfs, Neutrons Stars, and Black Holes

This module consists of three investigations, one at optical wavelengths, one at radio wavelengths, and one that makes use of the latest gravitational-wave data from the US's LIGO and EU's Virgo detectors.

## 6.1. Module 5A. Low- and High-Mass Star Death

During the next lunar dark cycle, students carry out another in-depth, astrophotography project at optical wavelengths, (1) of a protoplanetary nebula, (2) of a planetary nebula and its white-dwarf stellar remnant (see Figure 8), (3) of the stellar wind-blown bubble of a massive, Wolf-Rayet star, or (4) of a supernova remnant (SNR). More so than in Module 3, the emphasis is on narrowband imaging, but again with archival MIR layers (and in the case of the Crab SNR, with an archival NASA Chandra X-ray layer, which we provide). This module has both STEAM and physical-interpretation (temperature and density) components.

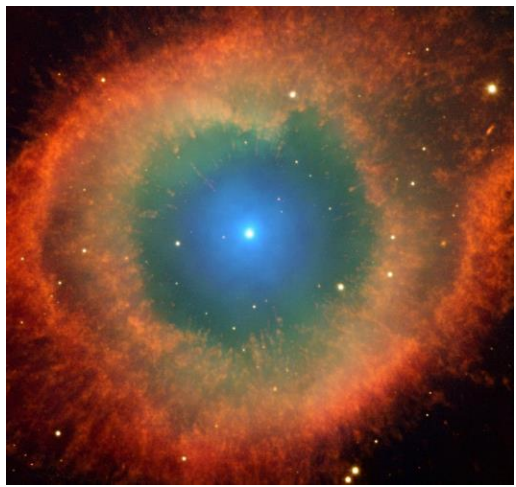

Figure 8. LRGB image of the Helix planetary nebula, and its white-dwarf stellar remnant, acquired using Skynet's 16-inch diameter PROMPT-6 telescope at CTIO. This has been supplemented with NASA Spitzer MIR data at 8 microns to reveal warm dust, and NASA WISE data at 22 microns to reveal cooler dust.

### 6.2. Module 5B. Pulsars and Polarization

One aspect of star death that students cannot study in Module 5A is neutron stars, which they can study at radio wavelengths in the form of pulsars. Each team (1) observes bright pulsar PSR 0329+54, (2) observes a handful of fainter, more challenging pulsars from a curated list, and (3) carries out a polarization calibration observation (a four-petal daisy map over a bright, mostly unpolarized source). Students show that PSR 0329+54 is polarized, and hence non-thermal, which they then connect to a model of the neutron star's rotation and magnetic field (see Figure 9). Students sonify and place size limits on (1) PSR 0329+54, (2) the fastest of the other pulsars that they observed, and (3) using archival data, PSR 1748-24, the fastest known pulsar.

### 6.3. Module 5C. Ripples in Spacetime: Gravitational-Wave Events

The other aspect of star death that students cannot study in Module 5A is black holes. However, black hole – black hole mergers are now detected fairly frequently at two US and one EU facility, called LIGO and Virgo, respectively. These instruments detect sub-nuclear scale ripples in the fabric of spacetime itself. We have built a tool that allows students to grab the official data products and model them themselves, measuring the masses and distance to merging objects, and more (see Figure 10). Students also (1) model LIGO data from binary neutron-star merger GW 170817, for which Skynet's PROMPT-5 telescope at CTIO co-discovered the first, and so far on-ly, optical counterpart to a gravitational-wave event, and (2) reproduce the counter-part discovery, using optical images of ≈40 galaxies collected that night.

Like Module 2, we believe that this module also has the potential to be used very broadly.

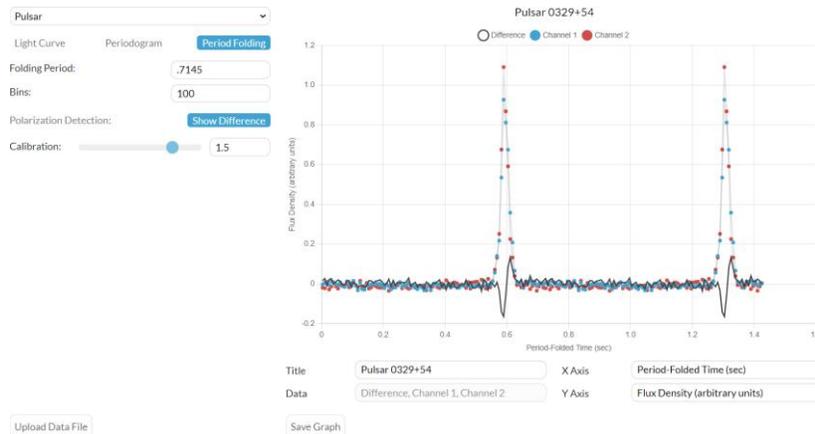

Figure 9.  Period-folding of one minute of data from bright, slow pulsar 0329+54, collected with Skynet's 20m-diameter radio telescope at GBO.  The non-zero difference in the orthogonal polarization channels shows that one channel's pulse leads the other's, demonstrating polarization and non-thermal emission (caused by electrons moving along the rotating neutron star's magnetic field lines).

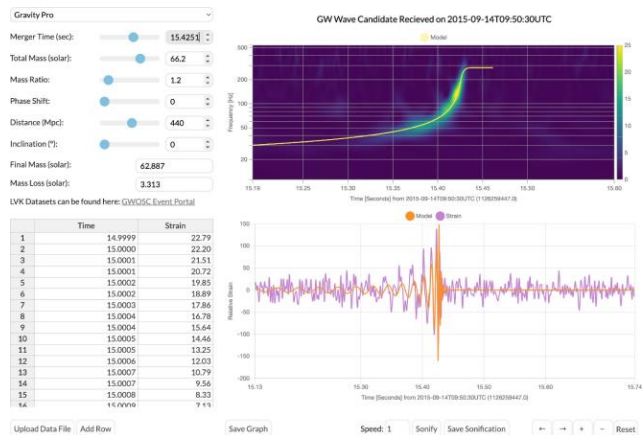

Figure 10.  Gravitymancer modeling of binary black-hole merger GW 150914.  The top image depicts frequency vs. time, and students use it to constrain the masses (and mass loss to gravitational waves) of the merging objects.  The bottom image depicts strain vs. time, and students use it to constrain the distance to (and phase of) the mer-ger.

## 7. Module 6. Radio Doppler Imaging and Spectroscopy: Weighing Galaxies

Students begin this module by mapping our sister galaxy, Andromeda. Being large and nearby, Andromeda is resolvable to our radio telescopes. However, they carry this out in high-resolution spectral mode, centered on the neutral hydrogen (HI), radio emission line. Then, as in Module 4B, they split this into RGB sub-bands, and visualize Andromeda's rotation in color (see Figure 11). They calculate Andromeda's line-of-sight velocity, collision timescale, rotational velocity, and enclosed mass.

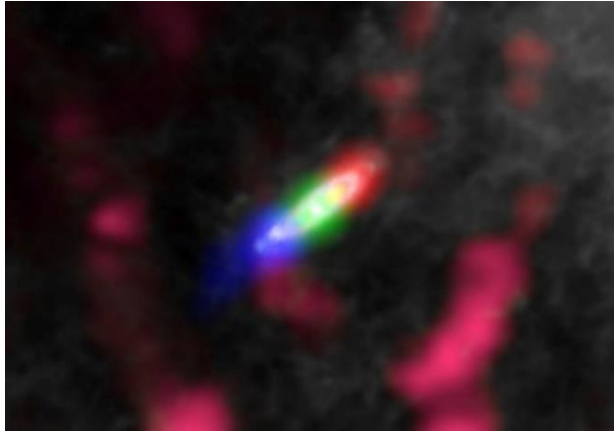

Figure 11. Doppler color image of Andromeda's HI emission line, acquired using Skynet's 20m-diameter radio telescope at GBO. (Pink is hydrogen in our galaxy, and gray is dust, from NASA IRAS FIR data.)

Then, students pick a farther out, but still relatively nearby, edge-on spiral galaxy from a curated list. Too far away to resolve with imaging, they instead acquire an "on/off", high-resolution spectrum, detecting the HI emission line (see Figure 12). From it, they calculate the galaxy's recessional velocity (and Hubble's constant), rotational velocity, and enclosed mass.

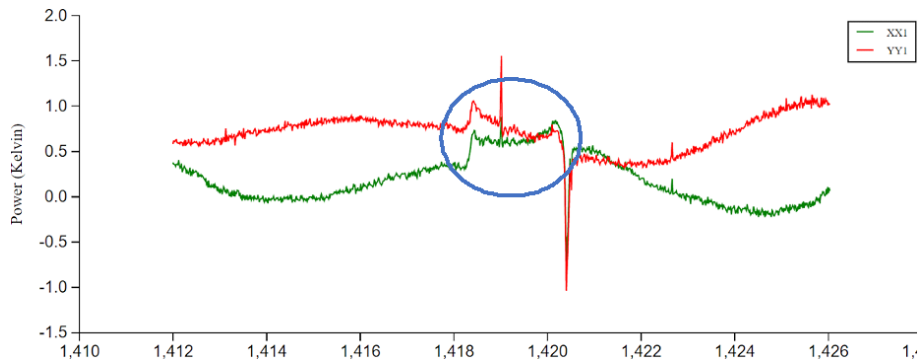

Figure 12. The same, hydrogen emission line, but from NGC 253, the Sculptor galaxy, made with Skynet's 20m-diameter radio telescope at GBO. Instead of appearing at 1420.4 MHz as it would in a lab, it is redshifted to lower frequencies, due to the expansion of the universe. Furthermore, it is broadened, due to the galaxy's rotation.

## 8.    Module 7.  Normal and Active Galaxies and their Evolution

During the final dark cycle of the semester, students carry out a capstone study, of a normal spiral, starburst, Seyfert, dwarf, normal elliptical/lenticular, and/or radio galaxy.  Students carry out LRGB + Hα narrowband imaging, and supplement these with archival MIR images, ideally from NASA's Spitzer spacecraft (see Figure 13).  Both Hα (star-heated gas) and MIR (star-heated dust) are tracers of star formation (Modules 3 & 5A).  So is blue light (Modules 2 & 3), and students learn different approaches for correcting their images for embedded dust.

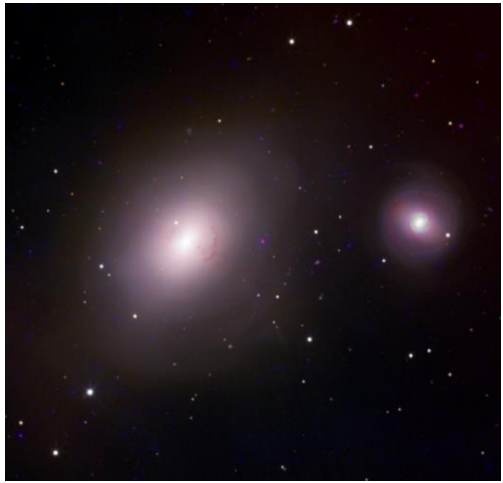

Figure 13.  LRGB image of lenticular radio galaxy Fornax A (left) and nearby spiral galaxy NGC 1317 (right), acquired using Skynet's 16-inch diameter PROMPT-6 telescope at CTIO.  This has been supplemented with NASA Spitzer MIR data at 8 and 24 microns, to reveal warm (pink) and cooler (blue) dust, respectively, both in these galaxies and in numerous background galaxies.

Simultaneously, students study their target at radio wavelengths.  They map their target, attempting to detect synchrotron emission from a central, supermassive black hole (Module 4A).  And they collect a high-resolution, on-off spectrum, attempting to detect HI (cold gas, also a tracer of star formation; Module 6).

Lastly, in an inter-team / class-wide activity, students organize their galaxies by their observed star-formation and synchrotron-emitting properties, using the following chart (Figure 14).  Although no one lives long enough to see galaxies evolve, they instead "connect the dots" of their case studies, revealing the two primary ways in which galaxies form and evolve.

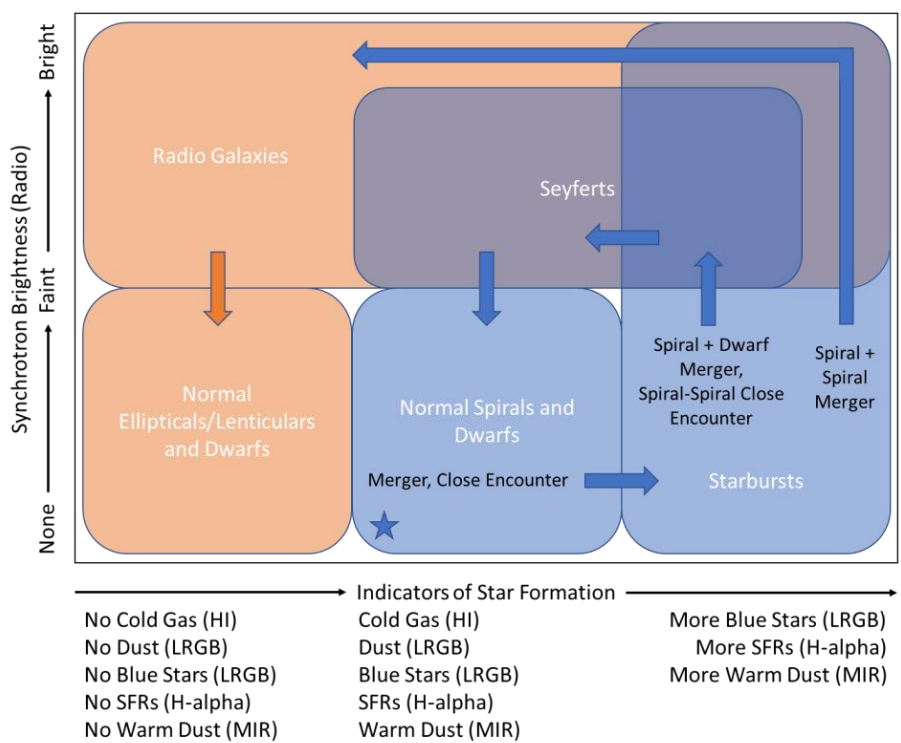

Figure 14. Students organize their case studies by their observed star-formation and synchrotron-emitting properties, which together then tell the story of galaxy formation and evolution (arrows).

### References

[1] Reichart, D. E. 2021, ASP Conference Series, 531, 233 (arXiv 2103.09895)

[2] Reichart, D. E. 2021, The Physics Teacher, 59, 728

[3] Reichart, D. E. 2022, ASP Conference Series, 533, 104 (arXiv 2202.09257)

[4] Reichart, D. E. 2023, ASP Conference Series, 537, 52 (arXiv 2304.02545)

[5] Janzen, D., et al. 2024, ASTROEDU2023: Astronomy Education: Bridging Research & Practice, Astronomy Education Journal, A110

[6] Fleenor, M., et al. 2024, ASTROEDU2023: Astronomy Education: Bridging Research & Practice, Astronomy Education Journal, A109

[7] Selph, L., et al. 2024, ASTROEDU2023: Astronomy Education: Bridging Research & Practice, Astronomy Education Journal, A101